# New Mechanism for Multiagent Extensible Negotiations

**Samir AKNINE**

LIRIS, Université Lyon 1

samir.aknine@univ-lyon1.fr

## ABSTRACT

Multiagent negotiation mechanisms advise original solutions to several problems for which usual problem solving methods are inappropriate. Mainly negotiation models are based on agents' interactions through messages. Agents interact in order to reach an agreement for solving a specific problem. In this work, we study a new variant of negotiations, which has not yet been addressed in existing works. This negotiation form is denoted extensible negotiation. In contrast with current negotiation models, this form of negotiation allows the agents to dynamically extend the set of items under negotiation. This facility gives more acceptable solutions for the agents in their negotiation. The advantage of enlarging the negotiation space is to certainly offer more facilities for the agents for reaching new agreements which would not have been obtained using usual negotiation methods. This paper presents the protocol and the strategies used by the agents to deal with such negotiations.

## Categories and Subject Descriptors

## General Terms

## 1. INTRODUCTION

Several recent studies on Internet users show the emergence of more and more sophisticated demands for these users in particular, for automated and flexible data processing systems. Among these systems, we can find those dedicated to automatic negotiation. One of the main research topics of the multiagent systems community focuses broadly on these problems. These systems which are inherently distributed focus generally on negotiation as a main means for solving collective and distributed problems of the agents. However the use of these systems in real applications on the Web still remains under investigation and experimentation.

Multiagent negotiation models rely on agent interactions through messages in order to solve specific problems. Although this research has already brought some solutions for agent negotiations, we should nevertheless point out that the use of current negotiation models in real world is still restricted. Indeed, these models are based on many assumptions and involve some constraints on the agent negotiation mechanisms. Unfortunately this makes use of automatic agent negotiations hardly convenient for. Additionally one important difficulty is to reproduce accurately to autonomous agents the behaviors of human negotiators in similar situations.

In this paper, we dealt with a new form of negotiation for multiagent systems which has not yet been addressed. This form of negotiation is denoted extensible negotiation. Unlike usual negotiations, this variant authorizes the dynamic extension of the set of items concerned with the negotiation in order to widen the space of possible solutions. This extension offers certainly more facilities for the agents to reach acceptable agreements which would not have been considered by current negotiation models. The prospect for such negotiations is to enhance convergence. Indeed negotiation convergence is currently considered as a hard problem in automatic negotiation systems.

In order to illustrate the complexity of such negotiations, and show their influence and impact, let us consider the example of an Internet user who wishes to organize his trip by the means of an electronic agency. The Internet user sends to this agency a query describing the list of services he wants to get in his trip, including plan tickets, a room reservation and museum tickets. This set of items is called the *support of the query*. The user also adds his list of preferences (*e.g.* dates of travel) and his constraints (*e.g.* hotel and room categories, budget, etc).

It is clear that at present few facilities are offered on the web to deal with such problems. Indeed, in current systems it is not allowed for users to negotiate with service providers. Current Internet systems proceed simply by querying existing databases to generate afterward proposals for the service applicants. The aim of our work is completely different since we intend, through the construction of these automatic negotiation systems, to allow the components of such systems to interact and negotiate in order to find a compromise satisfying both users and suppliers of services.

As for existing multiagent negotiation models and systems, they allow carrying out simple negotiations concerned with a single item or service [10,11,12]. There are also other models and systems dedicated to more complex negotiations which deal with combined negotiations and where sets of items or services are negotiated in parallel [2,3]. However these models do not allow to overcome the scope of earlier negotiations even if the context of the negotiation content concerns several items which could have different dependence relationships. Indeed, these models use as a reference the support of the query formulated at the beginning of the negotiation. As this support is limited to the items appearing in the user's query, the proposals formulated thereafter are restricted to this set of items which does not vary. Thus all the formulated proposals are necessarily related to one or more items in this query.

One limitation of this constraint appears, particularly, when the formulated proposals on this set of items are inappropriate to satisfy the requirements of the user. This results automatically in the failure of the negotiation without obtaining any agreement. In our model, we precisely look for ways to prevent these situations of failure. Agent negotiators can go beyond the limited set of items in the support of the query particularly when new compromising points between the applicant and the suppliers could be reached.

To illustrate the negotiation situations addressed in our model and for which current works fail to find a solution, let us take again our previous example on travel agencies. Once the query of the user (which includes in its support the following set of items: the plane tickets, the room reservation and the museum tickets) has been received and processed by the suppliers, if they do not have proposals to submit on this support, the failure of the negotiation will be imminent. The same situation also happens if the proposals on these items are not accepted by the applicant, for instance due to certain constraints (*e.g.* high cost); the failure of the negotiation will be also inevitable.

By means of the proposed negotiation model, we allow the openness of the negotiations with respect to the items which have not been mentioned in the support initially fixed by the user. Thus the negotiation started with the items: "plane tickets, room reservation and museum tickets" will be extended by the suppliers with new items, not initially specified in the query of the user, but which would interest him if they are related to the items listed in his query. On our example, the user could possibly agree to modify his constraints on his travel dates if the suppliers provide him additional services, *e.g. guided tours of historic sites*.

Although this service has not been initially specified by the user, the aspiration of the agency and the suppliers to conclude their negotiation with a winner-winner solution, can lead both of them to adjust its negotiation in order to satisfy each other. This is possible if each negotiator agrees to make some concessions on his own constraints in order to increase his utility compared to his initial state.

In this article, we tackle this problem of extensible negotiations and propose a new negotiation model which manages the constraints of this negotiation form. This model specifies the acceptable behaviors for the agents and respects consistency in the possible actions that might be carried out by the agents. A protocol describing these behaviors is proposed and deals with these different configurations.

This article is organized as follows. Section 2 described the problem and the context of extensible negotiations. Section 3 presents the negotiation model we propose and describes the protocol as well as the behaviors of each agent according to its role. The communication primitives allowed in the agent conversations and their semantics are also introduced. Section 4 illustrates then through an example the use of this protocol. Section 5 details the behaviors of the agents. Section 6 analyzes the properties of the protocol. Section 7 presents related work and section 8 is a conclusion on this work.

## 2. PROBLEM DESCRIPTION

To present the problem dealt with in this paper, let us consider again the example of the electronic travel agencies described previously. The role of these agencies is to handle the queries of the users by identifying the suppliers which would satisfy them. Each supplier can bid on one or more items of these queries. To model this situation, we use two types of agents: supplier agents and agencies. Agencies are represented by a set $A$ such that: $A= \{a_1, a_2, ..., a_n\}$ and the set of suppliers is denoted $F$ such that: $F= \{f_1, f_2, ..., f_m\}$.

The queries the agencies submit for the suppliers in $F$ are represented by the set $Q$, such that: $Q= \{q_1, q_2, ..., q_p\}$. Each $q_i$ considers a set of items in $S$ such that: $S= \{s_1, s_2, s_3, ..., s_r\}$. The constraints $C_i$ on items $s_j$ contained in each query $q_i$ (*e.g.* required dates) are indicated by each agency for its suppliers at the submission of the query.

## 3. NEGOCIATION MODEL

In this section, we formally define the concepts of extensible negotiation and then introduce the concepts we use in our model which allow the different agents to perform their interactions in this specific context.

### 3.1. Concepts of extensible negotiations

**Definition 1.** *Negotiation.* Let $X$ be a set of agents of a multiagent system $\Omega$. Let $O$ be a set of items in $\Omega$. Let $a_i$ an agent of $X$. Let $N$ be a negotiation on a set of items in $O$. Let $R$ be a set of roles for the agents of $X$ in $N$. $R$ contains at least two roles (supplier and agencies). Each agent $a_i$ in $X$ can participate in $N$ but with only one role of $R$ and several agents of $X$ may play the same role of $R$ in a negotiation. $N$ is a 6-uplet defined as follows:

$$N = (\Omega, X, O, R, f, C).$$

where:

- $f$ is a function defined on $X$ to $R$ and assigns for each agent in $X$ a role of $R$ in $N$.
- $C$ is a set of constraints defined either on the items under negotiation in $N$ or on negotiation (a limited duration for $N$, for instance). □

Each agent of $X$ owns a representation of the negotiation $N$ describing its negotiation process which references the different actions that this agent has performed during the negotiation. The negotiation process of an agent is thus defined as follows:

**Definition 2.** *Negotiation process.* Let $N$ be a negotiation on a set of items $o_j$ of $O$. Let $a_i$ an agent of $X$, $a_i$ plays the role $r$ of $R$ in $N$. The negotiation process of the agent $a_i$ in $N$, denoted $\Im_{ij}$, is defined as follows: $\Im_{ij} = (N, X^*, o_j, r, P, E, e_p, t_0)$, where:

- $X^*$ is the sub-set of agents in $X$ known by $a_i$ and involved in $N$. Each of these agents plays a particular role of $R$. $X^*$ contains initially the minimal sub-set of agents known by $a_i$ (if $a_i$ plays the supplier role, the size of this set is 1). This set is then refined by $a_i$ as the negotiation evolves since $a_i$ could be progressively informed about the other bidders making the same negotiation.
- $P$ is the negotiation protocol used by agent $a_i$ in $\Im_{ij}$.
- $E$ represents the set of ordered states for the negotiation $N$. This set draws the different proposals received (or sent) in order to guide the later strategic decisions of $a_i$.
- $e_p$ is the current state of the negotiation between the agent $a_i$ and the other agents of $X^*$.
- $t_0$ represents the beginning instant of the negotiation. □

As mentioned previously, the aim of our approach is to bring more flexibility for the negotiations by enlarging the space of the negotiated items between the agents.

**Property 1.** *Extensible negotiation process.* Let $N$ be a negotiation started at instant $t_0$ on a set of items $o_j$ in $O$. Let $a_i$ be an agent of $X$ with $a_i$ playing the role $r$ of $R$ in $N$. The negotiation process $\Im_{ij}$ of agent $a_i$ in $N$, is extensible, if $\xi(\Im_{ij}(o_j, t_0)) \neq \xi(\Im_{ij}(o_j, t_m))$ and $\xi(\Im_{ij}(o_j, t_0)) \cap \xi(\Im_{ij}(o_j, t_m)) \neq \varnothing$ at $t_o < t_m$ where $\xi(\Im_{ij}(o_j, t_k))$ defines the set of items $o_j$ at instant $t_k$ in the process $\Im_{ij}$. □

The set of items considered in the negotiation process varies during this specific time period due to the extensibility of the support of the query. To make a decision on a specific negotiation, an agent should measure its utility with respect to different possible extensions of the negotiation since a negotiation can be refined through time into several extensible negotiations.

Let us consider again our previous example, once an agency requested a set of items from its suppliers, each of these suppliers should be able to compare between different possible extensions of the negotiation in order to identify the one that will enable its to reach an acceptable compromise and gain a better utility. A supplier $â$ who would provide for instance, plane tickets, and room reservations should decide if it is necessary to extend its negotiation with a new set of items $o_1$ or $o_2$, where: $o_1$= {plane tickets, room reservation, guided tour} and $o_2$={plane tickets, room reservation, Opera tickets}.

**Definition 3.** *Dominance of extensible negotiations. Let N be a negotiation of an agent $a_i$, with $a_i$ playing the role r of R in N. Let $N_j$ and $N_k$ two extensible negotiations of N, respectively on two sets of items $o_j$ and $o_k$ of O. Let $U_i$ the utility function of $a_i$ in Ω. $N_j$ dominants $N_k$ if $U_i(N_j) > U_i(N_k) > U_i(N)$.* □

This definition shows that the utility of the agent $a_i$ in this new extensible negotiation should be higher than its utility in another extensible negotiation having a different set of items. This utility is measured according to the items concerned by the negotiation. Let us consider again our example, if the agency receives two different proposals on the previously defined sets $o_1$ and $o_2$, which one should it choose to support? The agency would be more interested by $o_2$ if the user prefers to make its visits without guides.

## 3.2. Negotiation protocol

The model of an extensible negotiation departs from usual negotiations by: (1) the dynamic changing of the items to be negotiated during a negotiation process as well as the constraints and the dependences concerning these items; (2) the need for synchronizing the behaviors of the agents in these negotiations since they concern several items; (3) the strategic behaviors which need to choose between several alternatives of the negotiation. This leaded us, particularly, in this part of work to define a suitable protocol for this form of negotiations. First, we give here the assumptions of the negotiation protocol we present below:

- *Information symmetry*. If an agency submits a query to a supplier, it knows that this agent proposes the required items. By the same way, if a supplier proposes new items which have not been explicitly specified in the query, it knows that the agency will be able to interpret these items and their relationship with the original query.
- *Completeness of the information*. At the begining of the negotiation, the query of an agency is completely specified, *i.e.* all the required items are explicitly indicated to the suppliers. The agencies do not intend to dissimulate information to the suppliers.
- *Limited duration for the negotiation*. Each agency and supplier agent knows, at the beginning of its negotiation, the time period allowed for the negotiation process. This duration may possibly be extended with the participants' agreements.
- *Penalization of the agents in case of decommitment*. Before that the negotiation starts the agents precise the moment after which the proposals and the answers given by each agent (*agency or supplier*) become committing. The commitments are explicitly specified by the semantics of the communication primitives sent by each agent during a negotiation. After this agreed instant, the decommitments of an agent will lead it to a penalization by the others. Here we do not discuss the value of this penalty. We assume that agents decide on it at the beginning of the negotiation. Recall that this assumption is often made by several protocols and negotiation systems [1,9,10].
- *Commitment symmetry and equity between agents*. No agent, supplier or agency, should be favored by the protocol. This will facilitate its acceptance by all agents. To do so, we allow each agent to decommit at least one time during its negotiation without being penalized. However, this decommitment should occur before reaching a certain state of the negotiation. The state where decommitments became unacceptable is known and agreed on in advance by all agents.

To carry out extensible negotiations, we propose a protocol based on several phases and iterations. In this protocol, agencies and supplier agents should be allowed to submit their queries in their initial form and to extend their negotiations afterward according to the evolution of their execution context. This allows the gradual extension of the negotiations beside the progression of the agreements between the agents.

### 3.2.1. Principles of the protocol

By defining this protocol, we want to facilitate its integration in real applications, such as the application of travel organization. We also search to guaranty the coherence and the flexibility of the agents' interactions since negotiations are carried out in an automatic way. Our protocol uses three phases: exploration, commitment and termination, represented in Figure 1.

- *Phase 1: Exploration*

This first phase starts when an agency sends the query of the Internet user to the suppliers. The agency specifies the requirements needed for the acceptance of the bids and waits for the reception of the proposals. The suppliers answer in this phase with their proposals, they detail their own conditions and they wait for the answers of the agency. At this stage of the negotiation, if the agents decommit they will not be penalized since their proposals are not committing them. The whole negotiation process is only delimited with a fixed duration but there is not a specific duration for this first phase of exploration.

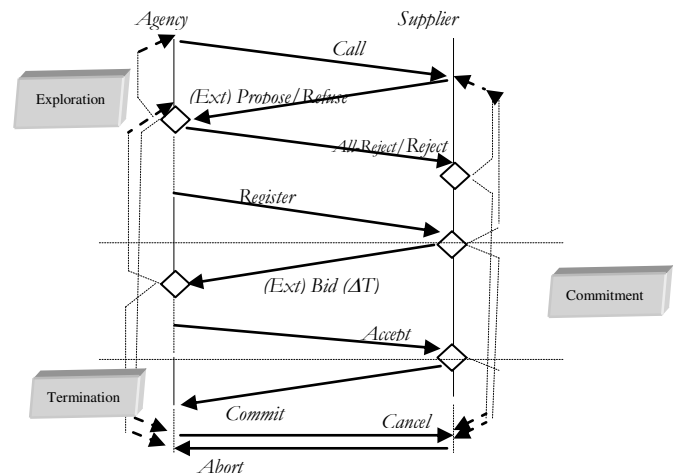

**Figure 1. Behaviors of the agency and the supplier agents in the negotiation.**

- *Phase 2: Commitment*

The beginning of the commitment phase is different for each proposal. This phase starts exactly when the agency accepts a proposal made by a supplier agent. After that if this supplier applies definitely, it should clarify the constraints on its proposal, in particular, its period of validity. Then this agent waits for the confirmation of the agency in the next period. If the agency desists before the end of this period, it will not be penalized. However if it desists after the expiration of the negotiation, it will be penalized. By the same way, a supplier which accepts the registration of an agency cannot cancel its proposal and if it does, even before the end of the validity period of the proposal, it will be penalized.

Among the constraints associated with a bid, we have the period of validity *(ΔT)* which indicates the moment after which this proposal could be cancelled automatically if the agency does no answer. Each agency should manage this kind of constraints on the bids. This constraint is certainly coherent, since for instance, currently to buy a plane ticket on the web, an Internet user can register for a proposal that interests it and adds an option on the validity period for this registration. The selling site is committed to provide the ticket as long as the validity period has not expired. The user should buy this item before the end of this period if it does not want to lose its ticket or to be penalized. Once this validity period expires, the supplier can, if necessary, propose new proposals of higher or lower values compared to the proposals made in preceding iterations.

*- Phase 3: Termination*

This phase starts once the agency confirmed to the supplier the acceptance of its proposal or its withdrawal from the negotiation. Any withdrawal of the agency must be performed before the end of the validity period of the bid; otherwise the agency will be penalized. After the agency accepts a bid, the acceptation of the supplier is expected. If a compromise between the participants has not been reached the negotiation fails.

### 3.2.2. Communication primitives of the protocol

Let us focus now on the communication primitives used by each agent during a negotiation process. These primitives concern a specific communication language, the semantics of which is given below. These primitives allow an interpretation of the exchanged messages between the agents. Let $K_f$ and $K_d$ be respectively the knowledge bases of the supplier agents *f* and the agencies *d*:

- *Cfp (d, f, Items, Conditions)*: The initiator of the negotiation, which is the agency *d* sends a message to the suppliers in which it declares its intention to negotiate the items described in its query. This query is completed with a set of specifications and constraints on the items. The pre and post conditions associated with this primitive are:

  *Pre-conditions(Cfp):*
  $\exists x \subseteq Items \land satisfied(x,d) \notin K_d \land wait(x) \notin K_d$

  *Post-conditions(Cfp):*
  $\exists x \subseteq Items \land wait(x) \in K_d \land satisfied(x,d) \notin K_f \land wait(x) \in K_f$

  - *wait(x)* means that the items in x are under negotiation.
  - *satisfied(x,d)* means that d owns a promising proposal for the set x.

- *Propose (f, d, Proposal, Cfp, Conditions)*: With this answer, the supplier *f* indicates for the agency *d* the proposals it can provide and for which the agency can register if it accepts them. Up till now *f* precises only the proposals it owns, it is still not committed with the agency.

  *Pre-conditions(Propose):*
  $\exists x \subseteq Items \land \exists y\ Proposal\ (x,y) \in K_f$
  $\land\ satisfied(x,d) \notin K_f \land know(y) \notin K_d \land wait(x) \in K_f$

  *Post-conditions(Propose):*
  $\exists x \subseteq Items \land \exists y Proposal(x,y) \in K_f \land y \in K_d \land sender(y,f) \in K_d$

- *Ext-Propose (d, f, Ext-Proposal, Cfp, Conditions)*. With this primitive, the supplier *f* indicates for the agency *d* that it has formulated a new proposal for the query it has made in the previous *Cfp*. However, the new proposal is an extended proposal since it contains new items that will probably interest the agency. Recall that the agency has not explicitly required them. It has thus to re-evaluate the new proposal taking into account this additional knowledge.

  *Pre-conditions(Ext-Propose):*
  $\exists x \subseteq Items \land \exists y\ Ext\text{-}Proposal\ (x,y) \in K_f$
  $\land\ satisfied(x,d) \notin K_f \land know(y) \notin K_d \land wait(x) \in K_f$

  *Post-conditions(Ext-Propose):*
  $\exists x \subseteq Items \land \exists y\ Ext\text{-}Proposal(x,y) \in K_f \land y \in K_d$
  $\land\ sender(y,f) \in K_d$

To build an extended proposal, the supplier agents compute the semantic distance between the items indicated in the user's query and the items they own in their domain ontology [8]. The constraints fixed by the user are also used to delimit the search space of possible proposals. This distance is used to identify the closest items to propose for the agents and which could interest the user with respect to its aspirations.

- *Refuse (f, d, Cfp)*: The supplier declares that it is not able to satisfy the original query made by the agency in its *Cfp*. It has no proposal to submit in this exploration phase.

  *Pre-conditions(Refuse):*
  $\exists x \subseteq Items \land satisfied(x,d) \notin K_f \land$
  $\exists y\ Proposal\ (x,y) \in K_f \land wait(x) \in K_f$

  *Post-conditions(Refuse):*
  $\exists x \subseteq Items \land wait(x) \notin K_f$

- *Register (d, f, (Ext)Proposal, Items, Conditions)*: With this answer, the agency *d* notifies to the supplier *f* that its proposal (possibly extended) is in a favorable position compared to all the received proposals either of *f* or other suppliers. The agency *d* may also provide more precise details on the conditions it would see improved by the supplier *f* in its bid. Up till now *d* is not committed with *f* even with this registration for its items. Both agents can withdraw from their negotiation in this stage of the negotiation without being penalized.

  *Pre-conditions(Register):*
  $\exists x \subseteq Items \land \exists y\ (Ext)Proposal(x,y) \in K_d \land sender(y,f) \in K_d$
  $\land\ satisfied(x,d) \in K_d \land preferred(x, d, f) \notin K_d$

  *Post-conditions(Register):*
  $\exists x \subseteq Items \land preferred(x,d,f) \in K_d \land satisfied(x,d,f) \in K_f$

  - *preferred(x,d,f)* denotes that the agency d considers the supplier f as the agent with the best proposal on x.
  - *satisfied(x,d,f)* signifies for f that its proposal on x interests d.

- *Reject (d, f, Items, (Ext)Proposal, Favored (Ext)Proposal)*: With this primitive, the agency *d* informs the supplier *f* that its proposal (possibly extended) is not the best preferred compared to those received from other suppliers. Here the proposal *Favored (Ext)Proposal* is the one likely to be accepted by the agency. *f* may thus improve its proposal within this negotiation by making some modifications on it (either in its extensions or not).

  *Pre-conditions(Reject):*
  $\exists x \subseteq Items \land \exists y\ (Ext)Proposal(x,y) \in K_d \land \exists z\ sender(y,z) \in K_d$

∧ z ≠ f ∧ satisfied(x,d)∈$K_d$ ∧ preferred(x, d, z)∈$K_d$

   Post-conditions(Reject):
   ∃x⊆Items∧ rejected(x,y,d)∈$K_f$∧ rejected(x,y,f)∈$K_d$

- *All-Reject (d, f, Items, Refused (Ext)Proposal, Best Refused (Ext)Proposal).* This message is sent by the agency and informs the suppliers that it has not selected a proposal among those received. It rejects all these agents and indicates for them the best refused proposal so as it allows them to improve their next proposals.
   Pre-conditions(All-Reject):
   ∃x⊆Items∧∃y Proposal(x,y)∈$K_d$ ∧ satisfied(x,d)∉$K_d$

   Post-conditions(All-Reject):
   ∃x⊆Items∧ ⌐∃y Proposal(x,y)∈$K_d$

- *Reject-and-new-Cfp (d, f, (Ext)Proposal, new Cfp).* The agency sends this message when all the proposals previously received are not compatible with its specified constraints on the required items. It should thus prepare a new call more adapted to the current context and send it to all the suppliers which have already answered with their proposals.
   Pre-conditions(Reject-and-new-Cfp):
   ∃x⊆Items∧ ∃yProposal(x,y)∈$K_d$ ∧ satisfied(x,d)∉$K_d$

   Post-conditions (Reject-and-new-Cfp):
   ∃x⊆Items ∧ satisfied(x,d)∉$K_f$ ∧ ⌐∃yProposal(x,y)∈$K_d$
   ∧ wait(x) ∈$K_f$

- *Bid (f, d, Cfp, Proposed Conditions)*: The supplier sends this message to the agency as an answer for its registration for the items. The difference between the proposal which it has submitted in the exploration phase and the current *Bid* lies only on the validity conditions added to this proposal. These conditions specify, the expiry time period of the proposal, for instance. The agency knows now, that it should commit (or give up its registration) before reaching the expiry limit of the bid. If it exceeds this limit, it will certainly lose its registration (and can be possibly penalized if the conditions agreed on authorize that). At this step, even if the agency carried out a registration for the items of the supplier *f*, it is not yet committed.
   Pre-conditions(Bid):
   ∃x⊆Items∧ satisfied(x,d,f)∈$K_f$ ∧Timeout(x)∉$K_f$
   ∧⌐∃y Bid(x,y)∉$K_d$∧sender(y,f)

   Post-conditions(Pre-Bid):
   ∃x⊆Items∧∃yBid(x,y)∈$K_d$∧sender(y,f)∧ committed(x,f,d)∈$K_d$

- Timeout(x) denotes the expiration of the decommitment phase.
- Committed(x,f,d) signifies that the bid on x has committed agent f to d.

The semantics of the primitive *Ext-Bid* is same than *Bid*, the only difference is in the set of items which became wider since the extension of the initial proposal.

- *Accept (d, f, Items, (Ext)Bid)*: With this message, the agency *d* accepts the *Bid* or the *Ext-Bid* of the supplier and it starts a termination phase for its negotiation with *f*.
   Pre-conditions(Accept):
   ∃x⊆ Items∧ ∃y(Ext)Bid(x,y)∈$K_d$ ∧ sender(y,f)∈$K_d$ ∧
   satisfied(x,d)∈$K_d$

   Post-conditions(Accept):
   ∃x⊆Items ∧ preferred(x,d,f)∈$K_d$ ∧satisfied(x,d,f)∈ $K_f$
   ∧ committed(x, f, d)∈$K_d$ ∧ committed(x, d, f)∈$K_f$

- *Commit (f, d, Cfp, (Ext)Bid)*: The supplier *f* sends this message when it receives an acceptance of the agency *d* on the bid the latter received previously and when the conditions of validity, such as validity period, are not violated. Once this message is sent, the supplier *f* is definitely committed and cannot decommit if it does not satisfy the conditions for a decommitment, for instance by accepting the payment of a penalty to the agency.
   Pre-conditions(Commit):
   ∃x⊆Items∧ committed(x,d,f)∈$K_f$∧
   ∃y (Ext)Bid(x,y)∈$K_d$∧ sender(y,f)

   Post-conditions(Commit):
   ∃x⊆Items∧∃y(Ext)Bid(x,y)∈$K_d$∧sender(y,f)∧
   committed(x,f,d)∈$K_f$

- *Abort (f, d, Cfp, (Ext)Bid, Reason)*: A supplier agent *f* can end its negotiation on a failure if for instance the validity period of a proposal has expired before receiving the answer of the agency. It thus sends a message to the agency explaining the reasons of the failure. This message ends the negotiation.
   Pre-conditions(Abort):
   ∃x⊆Items∧ committed(x,d,f)∉$K_f$∧
   ∃y(Ext)Bid(x,y)∈$K_d$∧sender(y,f)∧ ⌐Condition(y,d)

   Post-conditions(Abort):
   ∃x⊆Items∧ (Ext)Bid(x,y)∉$K_d$ ∧sender(y,f)∧
   committed(x,d,f)∉$K_f$∧ wait(x) ∉$K_f$

- Condition(y,d) signifies that the conditions on y have been respected by d.

Recall that in this method, we try to define a faithful protocol that guaranties coherent behaviors for the agents as done currently with human negotiations in the usual electronic systems (in plane reservation systems, for instance). The aim is certainly to facilitate use of this kind of protocols for the applications based on automatic negotiation.

## 4. ILLUSTRATIVE SCENARIO

Now, let's present an illustrative scenario of this protocol in order to show its use in a practical case and clarify its specification. Let us consider again the preceding example of the Internet user who intends to organize his trip. Let us suppose that this user submitted his query to an agency denoted Paul. Paul undertakes the negotiation of his items with two suppliers $f_1$ and $f_2$ able to satisfy this query. Paul's query considers plane tickets for New York for the period of July and a room reservation for 5 days. The budget for all these items is limited to $1000. At the initiation of the negotiation, *Paul* starts by making a call for proposals to $f_1$ and $f_2$ on the query *q*, where:
q= *plane tickets ∧ room reservation ∧ departure= Paris ∧ destination= New York ∧ stay= 5 dates∧ budget<=1000.*

To simplify, we denote the query sent by *Paul* to the two suppliers and the corresponding messages in the negotiation as follows:

   *Paul: Cfp —q→$f_1$ , $f_2$*

Each of the suppliers, after he has analyzed its available items formulates these proposals for *Paul*:
   $f_1$: Propose —Plane ticket ($700), Room($400)→ Paul
   $f_2$: Refuse —q→ Paul

*Paul* knows now that the supplier $f_1$ can provide him a plane ticket and a room reservation for $1100 and that $f_2$ can not answer its query. As an answer to these two proposals, *Paul* sends the two messages below:
   *Paul: Reject —Plane ticket ($700), Room ($400)→$f_1$*
   *Paul: Cancel —q→ $f_2$*

The first message means that the proposal of $f_1$ does not interest *Paul*, while the second message ends the negotiation with $f_2$. In the case that $f_1$ refuses to make another proposal, the negotiation is likely to be ended if *Paul* maintains its position in refusing the previous proposal. However if $f_1$ decides to modify its proposal, a new iteration of the negotiation will be started. In this case, the supplier $f_1$ which cannot lower its prices can decide to extend its proposal. It then sends the following message to *Paul*:

   $f_1$: Ext-Propose — Plane ticket, Room, Museum ($1100) → Paul

This extended proposal means that, for the same price the supplier $f_1$ agrees to add museum tickets to the preceding set of items. In case that *Paul* is still unsatisfied with this proposal, it can start again the series of rejections and proposals as long as the negotiation deadline has not been reached. Otherwise, *i.e.* if it is satisfied by this proposal, it sends the following message:

   Paul: Register — Plane ticket, Room, Museum ($1100) → $f_1$

This message indicates that *Paul* agrees with the proposal of the supplier $f_1$. Now, the supplier $f_1$ has to indicate the validity conditions for its proposal. To this end, it makes a complete bid for *Paul* as shown below:

   $f_1$: Ext-Bid — Plane ticket, Room, Museum ($1100) → Paul
   Constraints: Validity period=48h ∧ penalty=10%

The complete bid confirms the extended proposal of $f_1$ and that this proposal is valid only 48hours. If *Paul* does not make a final decision before this date, it will be penalized 10% on the total price. In the same way, $f_1$ is also committed with this bid that it should respect.
Now, in case that *Paul* refuses this bid, it is enough for it to send the following message before the end of the 48hours:

   Paul: Cancel — Plane ticket, Room, Museum ($1100) → $f_1$

This message enables it to cancel its registration with $f_1$ without being penalized. However, if it decides to validate its transaction with $f_1$, it sends the following message:

   Paul: Accept — Plane ticket, Room, Museum ($1100) → $f_1$

As the supplier $f_1$ has also committed in the previous phase, it has now only two possible choices of actions. Either it finishes the negotiation with *Paul* by sending the following message:

   $f_1$: Commit — Plane ticket, Room, Museum ($1100) → Paul

If it decides to cancel its proposal, it sends the message below, but agrees simultaneously to pay the penalty due its decommitment. The penalty will thus be paid in this case for *Paul*.

   $f_1$: Abort — Plane ticket, Room, Museum ($1100) → Paul

## 5. BEHAVIOURS OF THE AGENTS IN THE NEGOTIATION

To clarify our protocol, we detail the behaviors of the two roles of agents in the extensible negotiations. These behaviors are specified with Petri nets formalism.

**- Behaviors of the agency**
The protocol enables several series of exchanges for an agency with its different suppliers during the different phases of the protocol. Once the agency received the users' query, it contacts the suppliers it knows. Based on their proposals, the agency seeks to obtain the intended utility and to satisfy all the constraints on the items in the query. At the beginning, the agency makes a call for proposals, *Cfp*, to each supplier agent it considers as able to provide one or more required items (cf. Figure 2). Initially, supplier agents are not necessarily informed about the other supplier agents involved in the same negotiation.

The agency is now in the state 1 of the protocol where it waits for the answers of the contacted suppliers in the form of *Propose* messages. It waits until these suppliers compute their answers on the messages they have received. If these suppliers refuse to take part in the negotiation, for instance due to their impossibility to satisfy the conditions of the agency, the latter either ends its negotiation or makes some concessions (*i.e.* accepts to change the constraints on the items, for instance). Then the agency has to wait for the new answers of the suppliers. These two states are respectively involved by the transitions $Tr_5$ and $Tr_8$.

Once the suppliers have answered with their proposals, the agency goes to the state generated by the transition $Tr_4$ in order to process their messages. In this state, the agency analyzes the received proposals using its predefined strategy. Based on the received *Propose* messages and possibly extended proposals in the form of *(Ext)Propose*, it tries to build a solution which satisfies its constraints. If the agency succeeds, it starts its registrations with the *Register* messages it sends to its best supplier agents owning these proposals. It also sends *Pre-Reject* messages to the other suppliers. After this step the agency reaches the state 7 resulting from the $Tr_7$ transition. This step continues until the agency satisfies the query of the user.

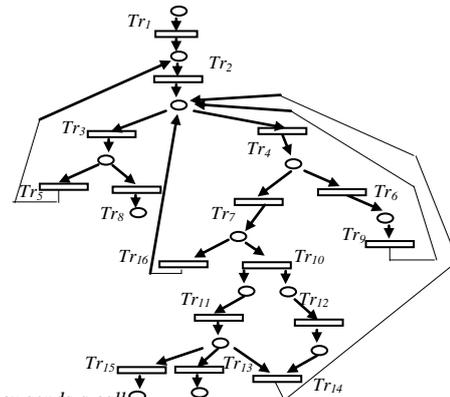

$Tr_1$: Agency sends a call
$Tr_2$: Agency waits the answers of the suppliers
$Tr_3$: Agency processes the refusals
$Tr_4$: Agency processes the proposals (possibly extended proposals)
$Tr_5$: Agency sends a new call
$Tr_6$: Agency rejects the suppliers
$Tr_7$: Agency registers with a potential supplier et pre-rejects others
$Tr_8$: Agency ends the negotiation on a failure
$Tr_9$: Agency waits the proposals
$Tr_{10}$: Agency waits the Bids or (Ext)Bid
$Tr_{11}$: Agency processes the Bids or (Ext)Bid
$Tr_{12}$: Agency looks for a new potential supplier
$Tr_{13}$: Agency ends the negotiation, and accepts a potential supplier and rejects the others
$Tr_{14}$: Agency modifies its potential supplier and informs the agents
$Tr_{15}$: Agency ends the negotiation with the agents on a failure
$Tr_{16}$: Agency receives new proposals

**Figure 2. State-transition machine of the agency**

Once the exploration phase finished, the agency waits for the final proposals of the potential suppliers. During the exploration phase, the agency can also receive temporary proposals of other suppliers it has only pre-rejected. These suppliers may indeed decide to make concessions and thus improve their previous proposals.
In this case, the agency reaches the next state following the transition $Tr_{16}$. In this state the agency processes its proposals based on its own strategies. It may decide to choose a new potential supplier agent for which it asks a final proposal and cancels its previous registration with the other potential suppliers. However after reaching a certain level of satisfaction of its utility function, the agency should only wait for final proposals from the potential suppliers. This is shown on the transition $Tr_{10}$ in the protocol description.

If a supplier agent has received a *Register* message and decided to decommit -or does not answer within a specific time period- the agency can prefer the transition $Tr_{11}$ where it selects a new potential supplier. However if the supplier agent answers with a final proposal in a *Bid* message or possibly an *(Ext)Bid* message, the agency analyzes the different allowed choices. If this *Bid* seems overestimated compared to the previous proposal received in the exploration phase, the agency may decide to definitively reject the potential supplier, or reject it temporarily, or to maintain it in its current state as long as the allowed negotiation time for this phase has not expired. Before rejecting the supplier, the agency makes sure that there is at least one other interested potential supplier having a better proposal. It should then send a registration message to this potential supplier for the items it proposes.

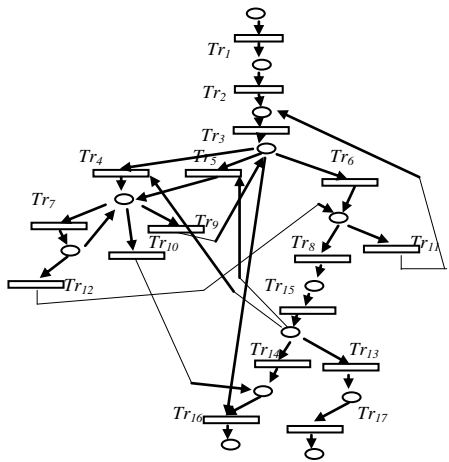

$Tr_1$: supplier receives a call from the agency
$Tr_2$: supplier prepares and sends a proposal
$Tr_3$: supplier waits for the answer
$Tr_4$: supplier receives a reject
$Tr_5$: supplier receives a collective reject (for it and the other suppliers)
$Tr_6$: supplier processes a registration for its items
$Tr_7$: supplier waits for the answer of the agency
$Tr_8$: supplier formulates a proposal in a Bid or (Ext)Bid
$Tr_9$: supplier formulates a different proposal
$Tr_{10}$: supplier formulates a new proposal in a Bid or (Ext)Bid
$Tr_{11}$: supplier prepares the termination of the negotiation
$Tr_{12}$: supplier receives a new registration
$Tr_{13}$: supplier receives an Accept message
$Tr_{14}$: supplier receives a Cancel message
$Tr_{15}$: supplier waits the decision on its Bid or (Ext)Bid
$Tr_{16}$: supplier ends the negotiation on a failure
$Tr_{17}$: supplier ends the negotiation on a success

**Figure 3. Behaviors of the supplier agent**

Once these messages sent, the agency waits for the final proposal. When it receives from a supplier a *Bid* or *(Ext)Bid* which satisfies it conditions, it sends an *Accept* message to this supplier where it announces that its proposal is definitively selected and that any negotiation concerning these items is closed. It waits for a *Commit* message from this supplier. This message results in the broadcast of the *Cancel* messages for all other suppliers. Thus the negotiation finishes successfully. However in case that the agency has been unsatisfied by all its suppliers, it ends the negotiation on a failure. This state results from the transition $Tr_{15}$ on the protocol description.

– **Behaviors of the supplier agent**

After the supplier agent received a call for proposals on its items, it analyzes this call in state 1 of the protocol (cf. Figure 3). In this state, the supplier prepares the proposal to give for the agency. It has to send a *Refuse* message if it considers that the query is not interesting or that the conditions of the query are unreachable. If it is able to meet these requirements, it sends its proposal in a *Propose* message and reaches then the state 3 where it waits for an answer from the agency on this message.

In this state, the supplier can receive a rejection message from the agency. It then behaves in different ways. If it builds a new proposal, it reaches the state 10. It can also decide to end the negotiation, and attains then the state 11. In this case, it waits temporarily until the agency revises some of its constraints and yields on some of its requirements. The supplier becomes temporarily accepted with a registration of the agency, this enables it to attend the state 6.

The negotiation stops in the case where a *Cancel* message is sent by the agency to the supplier. After the moment where a supplier agent receives a registration, it should formulate an answer in a *Bid* or *(ext.)Bid* message. This bid may be equal to its previous proposal formulated during the exploration phase or not. These behaviors are generated by two different transitions $Tr_8$ and $Tr_9$. The final proposal of the supplier agent makes it committed with the agency. Consequently, if the supplier receives a *Reject* message, either it returns to the previous state in order to wait for a new answer of the agency, or it decides to improve its proposal considering that other suppliers are on the same negotiation, and that it could lose the contract with the selected agency. Finally, if it receives an *Accept* message, the negotiation finishes and the supplier has only to finalize the transaction with a *Commit* message.

## 6. PROPERTIES OF THE PROTOCOL

The presented protocol supports negotiations between autonomous agents and guaranties several provable properties.

***Proposition 1****: The protocol is individually rational.*
A protocol is considered as individually rational if each agent finds an interest to adopt this protocol rather than to refuse it. It is enough to show that if the expected utility of an agent without using this protocol in negotiations is equal to 0, its expected utility will be higher than 0 using this protocol and in the worst cases remains equal to 0. The analysis of the different sequences of exchanged messages during the negotiation using this protocol can easily confirm this property since at each moment any agent is free to continue or stop its negotiations without any penalty except of course if it confirms its proposals.

***Proposition 2****: The protocol is robust to uncertainty in agents' proposals.*
In this case, we have to show that if an agent submits a proposal or an extended proposal (*i.e.* for a supplier agent) within a certain time period of the negotiation or if it accepts a proposal (*i.e.* for of an agency) this will not affect the other agents. The penalties are necessary at this negotiation level in order to manage the relative uncertainty in the agents' answers. Indeed, the role of the protocol is precisely to modulate the value of this penalty considering the uncertain reactions of each agent. It is thus enough that each agent detects the variations in the proposals of other agents and consequently increases or decreases the value of the penalty.

***Proposition 3****: The negotiation between agents using this protocol ends necessarily after a finite number of steps.*

For that, it is enough to show that there is no infinite loop in the behaviors of the agents during the phases of submitting proposals, acceptance or refusal of proposals. This could be shown by successively tracking the sequences of exchanged messages between the agents in the different phases.

## 7. RELATED WORK

Multiagent negotiation study is inherently interdisciplinary. Several techniques have been used to design innovative mechanisms for complex negotiations. To tackle computational complexity, several authors proposed simultaneous auctions where bids are placed for individual items. [7] propose a model for simultaneous negotiation in forward and reverse auctions. They consider a service composition agent that both buys individual items by participating in many English auctions and sells composite services through Reverse-for-Quote auctions. The combinatorial complexity is tackled through the composition of items sold as a unique item. The drawback of this model is the risk taken by the seller which sells a composite service before having bought the components through English auctions.

[4] studied simultaneous ascending auction (SAA) where bids are placed for individual items rather than package of items. This auction model, as the previous one, reduces complexity and, as pointed out by the authors, is an effective method of auctioning many items. Simultaneous sales and ascending bids enable price discovery, which helps bidders, build desirable package items. [5] proposes to allow only a limited number of combinations of items. This model is similar to the previous ones where bids are done for predefined packages proposed as unique items.

[2] propose simultaneous clock auctions avoiding computational complexity. With multiple related goods, price discovery is important to simplify the bidder's decision problem and to facilitate the revelation of the preference in the bids.

Another stream of research addresses computational complexity through iterative auctions. [6] proposes *iBundle*, an *iterative* combinatorial auction in which agents can bid for combinations of items and adjust their bids in response to bids from other agents. *iBundle* computes the efficient allocation when agents follow myopic best-response bidding strategies, bidding for the bundle(s) that maximize their surplus taking the current prices as fixed. *iBundle* solves problems without complete information from agents and terminates in competitive equilibrium. Moreover, an agent can follow a myopic best-response strategy with approximate values on bundles.

Compared to this work and as previously showed, in our approach, we are also interested in combined negotiations but we certainly address them in a completely different way. Indeed we assume that our agents can submit proposals on items which are not mentioned in the initial specification of the negotiated items. This behavior seems completely acceptable since it can lead more easily, as showed in this article, to agreements when agents have different aspirations and when these agreements are difficultly reachable with restricted proposals. This negotiation model is original and takes really into account practical considerations of the new negotiation applications. Thus we have enriched the theory of negotiation in multiagent systems with a new practical model which can be also adapted to various contexts of use and not only to travel organization as shown on our examples.

## 8. CONCLUSION AND PERSPECTIVES

In this paper, we have proposed a new negotiation mechanism for multiagent systems. At present, several negotiation models have already been defined to deal with different issues in agent negotiations. Among these models, several tackled the negotiation problem either of only one item, or of several units of the same item or of several items. All these works are discussed in [2,3,8]. However, these mechanisms did not allow yet carrying out extensible negotiations.

The concept of extensible negotiation is introduced in this paper to take precisely into account the specificities of combined negotiations, such as dependence relationships among negotiated items, their impact on the negotiation mechanism and on the strategies, as well as the huge difficulty for reaching acceptable compromising points among agents. For this reason, we have proposed a new negotiation protocol dealing with these requirements. This protocol is based on several negotiation phases. We, then, have specified the communication primitives used by the agents and their semantics. Several properties of the mechanism have been studied. This work has also been extended to address other forms of dependences among the items, particularly items with different preference levels, priorities, etc. New protocols dealing with this kind of queries are still under study.

Additionally, a first implementation of this protocol has been performed to study the behaviors of the agents on a real application. The case studied is the example of travel organization. This implementation has particularly helped testing and understanding the impact of the strategies and behaviors of the agents and to identify the most relevant ones, in particular for accelerating the convergence of the agent negotiations.